# Revisiting Maya Blue and Designing Hybrid Pigments by Archaeomimetism**

*Catherine Dejoie, Eric Dooryhée, Pauline Martinetto\*, Sylvie Blanc, Patrice Bordat, Ross Brown, Florence Porcher, Manolo Sanchez del Rio, Pierre Strobel, Michel Anne, Elsa Van Elslande, Philippe Walter*

The search for durable dyes led several past civilizations to develop artificial pigments. Maya Blue (MB), manufactured in pre-Columbian Mesoamerica[1], is one of the best known examples of an organic-inorganic hybrid material. Its durability is due to a unique association after heating together indigo and palygorskite, a particular fibrous clay occurring in Yucatan. Despite 50 years of sustained interest, the microscopic structure of MB and its relation to durability remain open questions[2-7]. Combining new thermogravimetric analysis and synchrotron X-ray diffraction data with molecular modelling, we deduce a new explanation of the chemical stability and the durability of MB: steric shielding of the alkene bond and plugging of the clay channel entrances by very strongly attached indigo molecules. Elucidating the properties of historical and even laboratory-reconstituted MB offers the possibility to create new materials by archaeomimetism: engineering an archaeoinspired pigment, here indigo in a zeolite host, which satisfactorily reproduces the colour and chemical stability of MB. Comparing and contrasting ancient and modern MB and the archaeoinspired material confirms that chemical stability in MB depends more on steric shielding than on the strength of chemical bonding as currently accepted[8-10]. Archaeomimetism opens the way to new functional materials, combining present day chemical know-how with insights and inspiration from successful materials in our cultural heritage.

The Mayas invented a remarkable hybrid material by "mineralizing" the añil organic dye in palygorskite *via* moderate heating[2,11]. Resistance of the blue colour to time, chemical and physical degradation[1-2] (figs. 1, 2a) attests the exceptional stability of indigo in palygorskite. Colour fixing in archaeological MB predominantly involves one molecular indigoid species, absorbing at 660nm and fluorescing at 750nm (figs. 1, 2a).

[*] Dr. C. Dejoie, Dr. E. Dooryhée, Dr. P. Martinetto, Dr. P. Strobel, Dr. M. Anne
Institut Néel / CNRS-UJF
25 avenue des Martyrs, BP 166, F-38042 Grenoble Cedex 9, France
Fax: (+) 33 (0)4 76 88 10 38
E-mail: pauline.martinetto@grenoble.cnrs.fr

Dr. S. Blanc, Dr. P. Bordat, Dr. R. Brown
Institut Pluridisciplinaire de Recherche sur l'Environnement et les Matériaux, CNRS
2 avenue Pierre Angot, F-64053 Pau Cedex 9, France

Dr. F. Porcher
Laboratoire de Cristallographie, Résonance Magnétique et Modélisation, UHP-CNRS
Faculté des Sciences BP 70239 , F- 54506 Vandoeuvre-les-Nancy, France

Dr. M. Sánchez del Río
European Synchrotron Radiation Facility
Polygone Scientifique Louis Néel, 6 rue Jules Horowitz, F-38000 Grenoble, France

E. Van Elslande, Dr. Ph. Walter
Centre de Recherche et de Restauration des Musées de France, CNRS
Palais du Louvre, Porte des Lions, 14 Quai François Mitterrand
F-75001 Paris, France

Dr. C. Dejoie
Lawrence Berkeley National Laboratory - ALS
1 Cyclotron Road, CA 94720 Berkeley, USA

Dr. E. Dooryhée
Brookhaven National Laboratory – NSLS II
NY 11973 Upton, USA

Dr. F. Porcher
Laboratoire Léon Brillouin, CEA-CNRS
F-91191 Gif-sur-Yvette, France

[**] Synchrotron diffraction measurements at the ESRF-BM02 beamline benefited from the support of J-F. Bérar, N Boudet, S. Arnaud and B. Caillot. J. Kreisel provided assistance for the *in situ* fluorescence studies at the LMGP, Grenoble. P. Odier (Institut Néel) helped with the thermal analyses and P. Bordet with the single crystal X-ray diffraction studies. C.D. acknowledges CIBLE and MACODEV grants from Région Rhône-Alpes. Molecular Dynamics calculations were carried out at the MCIA computer facility, Bordeaux . This work was supported by A.N.R. grant CIS-007-005 (NOSSI project).

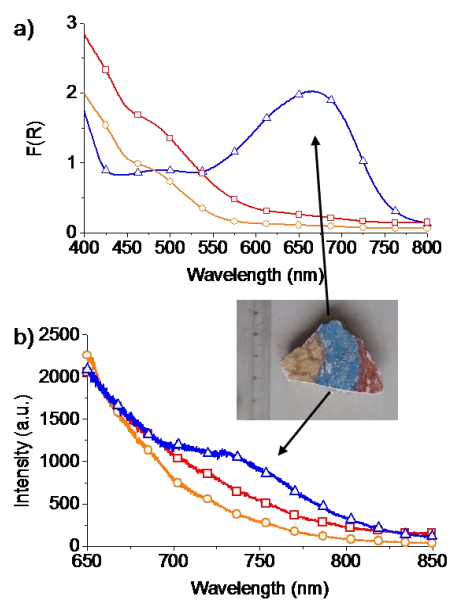

*Figure 1.* Reflectance, *(a),* and fluorescence spectra, *(b),* of a fresco fragment from Cacaxtla (Mexico). (○) yellow part; (□) red; (Δ) blue (Maya blue, MB). The main absorption band of the blue part is at 660nm. Strong emission from all parts at *ca.* 550 nm, might be due to the substrate. The blue part shows an additional weak fluorescence band at *ca.* 750 nm.

Comparison of the spectra of indigo in various hosts (fig. 2 and supporting information, SI fig. 1), including reconstituted MB (fig. 2a) and in solution [12] (SI fig. 2), indicates that the 660nm reflectance band in MB is due to monomeric indigo. Absolute wavelengths (fig. 2) are determined by interaction with the environment (*e.g.* a *ca.* 50nm red shift from MFI to more polar palygorskite)[13-14].

However, the location of indigo in MB remains controversial. According to Van Olphen[2] and Chiari *et al.* [15], the molecule preferentially lies in grooves at the surface of the clay fibres. On the other hand, Kleber *et al.*[3], Chiari *et al.*[5], and Fois *et al.* [16] assumed that indigo enters the internal channels, replacing zeolite water. In a third model proposed by Hubbard *et al.*[7], and favoured by Sánchez del Río *et al.*[17], indigo does not enter the clay but stays at the channel entrances.

internal channels. Chiari *et al.* [15], previously concluded that indigo does not replace any great quantity of channel water, and called for the "surface groove model" first proposed by Van Olphen[2]. Furthermore, heating bare palygorskite induces the departure of water and clay folding[18], with a modified diffraction pattern. Here, the diffraction pattern of laboratory-made MB with high enough levels of indigo (>5% wt.) is insensitive to temperature up to 250 °C, meaning that folding is hindered in the presence of indigo (SI fig. 3). Prevention of folding due to the presence of indigo molecules was reported recently by Ovarlez *et al.*, for an indigo@sepiolite hybrid [19]. To our knowledge, this is the first time that this aspect is shown for palygorskite. The location of indigo in MB thus affects the overall structure and rehydration of the clay, excluding that MB be an exclusively surface compound[7, 17].

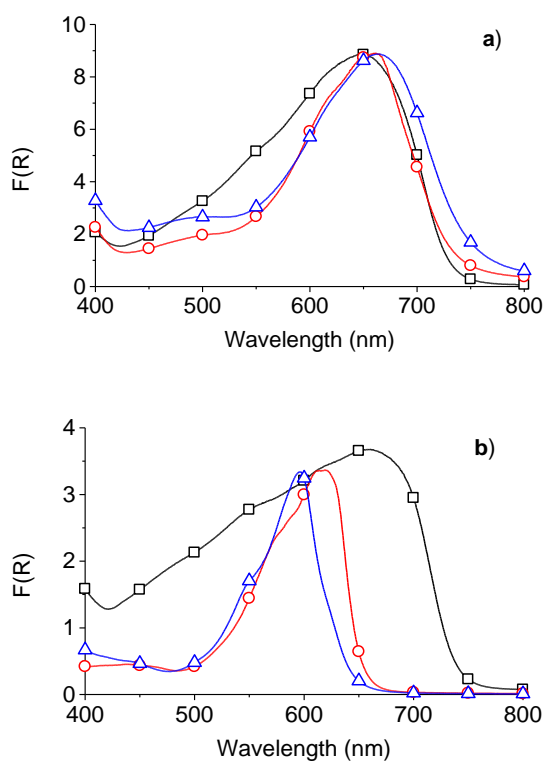

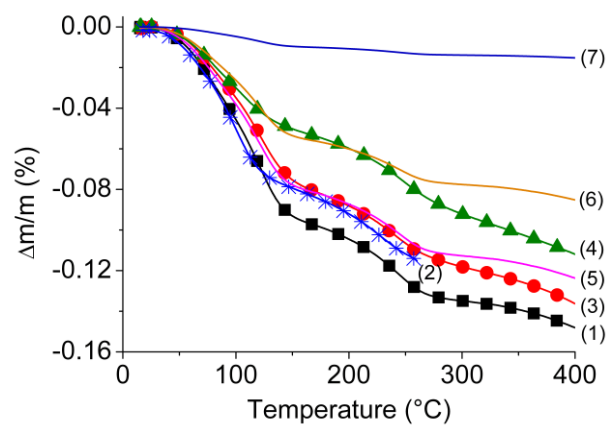

*Figure 3.* Points: experimental TGA data, 1) raw palygorskite, in agreement with ref [18] (■); 2) an indigo (10%wt)- palygoskite hybrid during formation (*); 3) 2%wt (●) and 4) 10%wt (▲) hybrids after rehydration. *Curves:* deduced from data *1)*, assuming indigo replaces internal channel water, for: *5)* 2%wt indigo; *6)* 5% and *7)* 10% (theoretical maximum 11%).

*Figure 2.* Kubelka-Munk transforms, F(R), of the UV-visible diffuse reflectance of indigo-microporous guest-host systems (2%wt. Indigo): *(a)* indigo-palygorskite (*i.e.* reconstituted MB); *(b)* indigo-silicalite. Symbols: □ unheated mixtures; ○ heated hybrids; Δ heated hybrids after the nitric acid test. The absorption band responsible for the blue colour persists after the nitric acid test. Spectra are normalised at 665 nm (a) and 610 nm (b).

In trying to clarify this issue, we performed new *in situ* and post-formation X-ray diffraction (XRD) and thermogravimetric (TGA) experiments on MB. Figure 3 shows TGA data *in situ* and after rehydration. The solid line curves are simulated TGA curves for rehydrated material, deduced from that of raw palygorskite on the assumption that indigo replaces mobile channel water from the host (SI section 3). Figure 3 indicates that *i)* indigo does not impede the escape of channel water from palygorskite during hybrid formation (onset at *ca* 150°C); *ii)* rehydration at room temperature is significant; *iii)* at high loading, indigo only partially lies in the

Summing up earlier and our new data, part of the indigo in MB is within the channels, part on the surface. But why are the channels not filled to capacity at high loading, why is heating required to produce a stable hybrid, why is MB so stable on return to room temperature?

We performed molecular dynamics optimizations and simulations at finite temperatures to improve our understanding. Contrary to the three dimensional periodicity of models in earlier studies[5,8,10,16], this is to our knowledge, the first simulation of indigo adsorbed on a *free-standing* palygorskite fibre, allowing comparison of internal channel, wide and narrow surface grooves and channel entrance sites. We prepared a model fibre prism, about 30 Å x 40 Å x 50 Å in the *a,b* and *c* crystal directions (*ca.* 11200 atoms), from the crystal data of the monoclinic (room temperature) phase[18], retaining all cavity and structural water. Dangling oxygens on the external surfaces were capped to form hydroxyl groups. Simulations of un-doped fibres and bulk palygorskite agreed well with the data of ref [18]. We added test indigo molecules in four locations: *(1)* at the fibre tip, *(2)* in an internal channel and *(3)* and *(4),* in the surface grooves of the (100) and (010) faces. The simulations predict that the most stable sites for indigo are, in order: *(1)* at the channel ends, astride the siloxane bridges between the octahedral sheets and thus partially obstructing the channels (adsorption energy $E_1$=-22 kcal/mol, fig. 4a); *(2)* within the bulk channels ($E_2$=-7.7 kcal/mol); *(3)* in the wide grooves of the (100) face, ( $E_3$=-6.6 kcal/mol) and *(4)* in the narrow grooves of the (010) face ($E_4$=-3.1 kcal/mol). Site *(3)* is discarded below since the (100) face is not an expected preferred crystal growth plane[20].

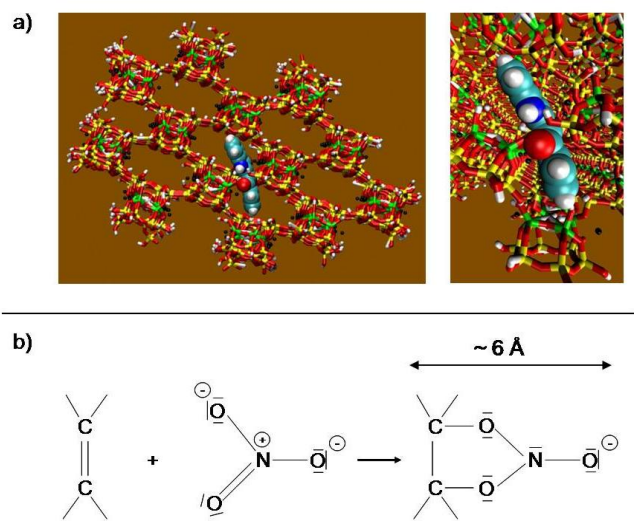

***Figure 4.*** *(a)* Molecular model of the most stable adsorption site of indigo in MB, at the channel ends (partial orthographic and perspective views down *c*, with water withdrawn for clarity). *(b)* Oxidation of the alkene function of indigo by nitric acid proceeds via a bulky, bridging intermediate that must lie perpendicular to the molecular plane of indigo.

Now from the size of the indigo molecule, palygorskite crystal data, and the typical size and shape of the fibres[20], we estimate the maximum possible loadings for the fibre tips, the surface grooves and the channels in the bulk to be around 0.05, 0.5 and 11% (% mass out of total). Combining these estimates with the experimental XRD, spectroscopic and TGA data and the simulations, we can now offer a new understanding of the formation and temporal stability of MB.

On warming indigo and the clay together, indigo first populates the very stable site at the fibre tip (1) and the groove sites (4) (exothermic processes). On steric and energetics grounds, 'doorkeeper' molecules in site *(1)* allow escape of cavity water but oppose population of the internal channels *(2)* until at higher temperatures (barrier $E_2$-$E_1$) they are activated further into the channel entrances. Diffusion within the channels involves an inherently slow process, single file rather than regular diffusion, so only partial filling may result from insufficient baking. Randomly diffusing indigo in the channels prevents the clay from folding at higher temperatures, maintaining the XRD signature of the low temperature phase (SI fig. 3). On cooling, water reoccupies remaining free channel volume, but once indigo is locked within the channels, by the 'doorkeeper' molecules, the barrier to egress (furthermore at room temperature), is higher, at least $E_4$-$E_1$ to dislodge a doorkeeper. This accounts for the historical stability of MB. Indigo in the channel entrances thus acts as its own thermally activated one-way valve or diode.

While exact site energies will depend somewhat on the level of theoretical chemistry in the modelling, we may propose a new general picture of MB, with monomer indigo indeed partly within the channels as proposed first by Kleber *et al.*[3], where it prevents folding, indeed partly in external surface grooves[2] and indeed partly inserted into the channel openings of the clay substrate, as proposed by Hubbard et al.[7] We view the last category as a very minor component, but crucial to the formation and stability of MB.

Knowing the location of the indigo, we now address the chemical stability of MB, particularly under oxidising conditions. While transformation of blue indigo into yellow isatin in nitric acid involves breaking of the central C=C double bond of the former, attack of the alkene function requires room for a perpendicular bridging intermediate (Fig. 4b) [21]. In sites *(1)*, *(2)* and *(4)*, the local environment indeed screens the alkene function of indigo from oxidation since the reaction intermediate cannot be accommodated. We thus believe that fixing of the indigo molecule to sterically screened sites mainly accounts for the chemical stability of MB.

Keeping in mind our interpretation of the structure of MB, and its remarkable chemical stability by the shielding of the indigo alkene function in specific sites, we applied this knowledge to the design of new, archaeoinspired hybrid pigments.

Copying the general channel structure of palygorskite and looking for screened sites for indigo, we examined zeolites as possible microporous hosts. We chose the hydrophobic, high-silica zeolite MFI (silicalite) [22], with channel section (5.3x5.6 Å²) slightly larger than the width of the indigo molecule (4.8 Å). Preparation was similar to the procedure for reconstitution of MB (experimental section). Figure 2 shows the reflectance bands of indigo in palygorskite and silicalite hosts before and after baking and after the oxidising test. Indigo and the zeolite MFI give rise to a new hybrid pigment successfully enduring exposure to concentrated nitric acid (Gettens stability test)[1]. Colour stability is attested by the persistence of the diffuse reflectance band at 600nm (fig. 2b). The colour of the archaeomimetic pigment is highly light- and heat-resistant (SI fig. 4). Crystallographic positions of the indigo molecules were investigated combining single crystal and powder XRD. Our diffraction Fourier syntheses reveal extra electron density associated with indigo molecules in two sites in the channel structure (fig. 5). Indigo in the silicalite network predominantly lies at the channel intersect, with molecules partly engaged in the sinusoidal channels. Molecules occupying the second site at high loading lie longitudinally in the straight channels.

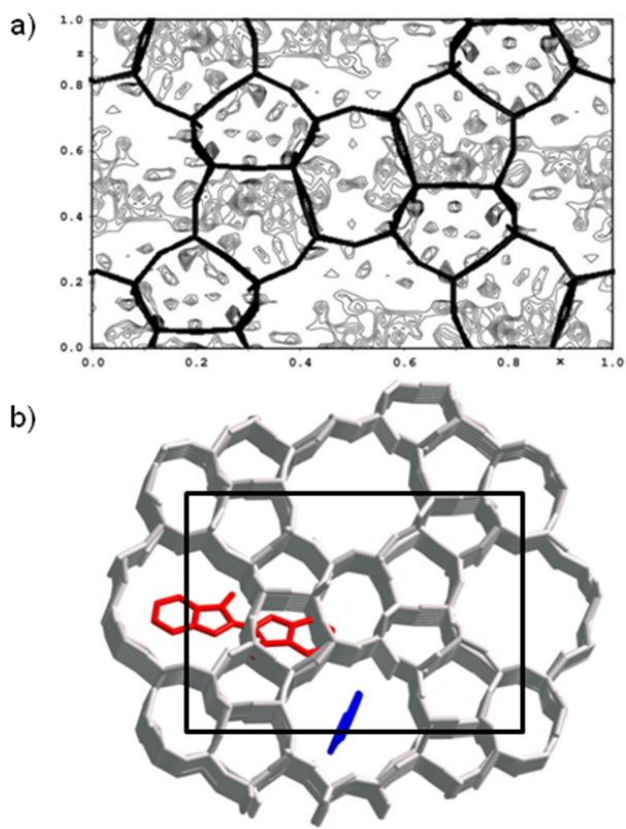

***Figure 5.*** Indigo@MFI (5%wt, 300°C): *(a)* Single crystal Fourier difference map with extra electron density associated with indigo molecules lying at the channel intersect and longitudinally in the straight channels; *(b)* Crystallographic positions of the indigo molecules in the channel structure of silicalite. Projections along the straight channel axis. For clarity, only one molecule of each site is drawn.

Considering the channel dimensions of the MFI zeolite and the positions of the dye, the reaction intermediate inducing the transformation of indigo into isatin cannot be accommodated (fig. 4b). Thus, steric considerations alone suffice to explain the chemical stability of the indigo@MFI archaeomimetic pigment. This aspect is confirmed by the failure of hybrids based on LTA zeolite (channel section 4.1x4.1 Å²) and mordenite (channel section 7x6.5 Å²) to pass the Gettens test (SI fig. 1). Furthermore, since both aluminium and internal water are absent in high-silica silicalite (MFI), but abundant in these other hosts (see SI), the particular position of the dye in MFI zeolite appears more important than possible complexation by Al or hydrogen-bonding to water. These results confirm our conclusion on the chemical stability of the archaeological pigment MB.

This comparison of ancient and modern hybrid pigments provides a first example of how study of archaeo-material may improve our understanding of the properties of functional materials[23]. The present example inspired by MB leads to a non-toxic, acid-proof, low-cost and stable hybrid pigment. Archaeomimetism undoubtedly opens a wide range of future studies concerning materials and properties with counterparts in our cultural heritage.

## *Experimental Section*

Archaeological samples of mural frescos from the Cacaxtla archaeological site (Mexico) were provided by C. Reyes-Valerio. MFI zeolite samples were prepared using an adaptation of the fluoride route to hydrothermal synthesis[24] using tetrapropylammonium cation as template. Samples were calcined at 600°C in order to remove the organic template. The palygorskite clay (also called attapulgite) came from Ticul in the Yucatán peninsula (Mexico), a major source of clay for Maya Blue for more than eight centuries. Synthetic indigo (Sigma-Aldrich) was used.

To prepare zeolite hybrids, the inorganic matrices and indigo (0.5, 1, 2, 5, 10% wt.) were hand-ground and mixed in a mortar. The resulting fine powders were pressed into a pellet to enhance contact between the two components. The pellets were baked in air in an oven for 5 hours at different temperatures 100°C, 150°C, 200°C, 250°C or 300°C (data shown here). After the heating phase, pellets were re-ground and washed with acetone before being exposed to nitric acid at room temperature to test the stability of the materials. A few milligrams of the powder was stirred for 10 min in concentrated $HNO_3$ (14M). Stability of the sample was attested by the persistence of the blue colour.

Diffuse reflectance spectra were recorded on a Varian Cary 5 spectrometer equipped with an integrating sphere. The spectra are displayed as $F(R)$ Kubelka–Munk units, with: $F(R) = (1-R)^2/2R = k/S = \varepsilon c/S$ where $R$ is the corrected reflectance, $S$ stands for the scattering coefficient (depending on the size and form of the particles), $\varepsilon$ the molar absorption coefficient of the analyte and $c$ its molar concentration. Corrected fluorescence spectra were measured at room temperature on an Edinburgh FSL900 spectrofluorimeter. TGA and DSC were conducted on a SETARAM TGA 24 thermoanalyser. Samples (~15 mg) were heated under nitrogen from 25 to 700 °C, at a rate of 10°C/min. For powder XRD measurements, the powders were hand-packed in 1mm glass capillary tubes and mounted on the goniometer head of the 7-circles diffractometer of beamline BM02-ESRF[25]. Powder diffraction (XRD) data were collected in the high-resolution 2θ step scanning mode, and data were processed using the Fullprof software[26]. Single crystal data were collected on a KappaCCD diffractometer (Bruker-Nonius) and refinements were performed with the single-crystal package JANA[27].

We used dl_poly 2.20[28] for molecular simulations, with the force field of M. Matsui for the palygorskite host[29], the SPC/E model of water[30] and the OPLS-AA force field for indigo[31] and the Lorentz-Berthelot rules for cross terms. Electrostatic interactions were calculated with either Ewald summation or electrostatic damping[32].

[1] R.J. Gettens, *Am. Antiquity* **1962**, *7(4)*, 557-564.
[2] H. Van Olphen, *Science* **1966**, *154*, 645-646.
[3] R. Kleber, L. Masschelein-Kleiner, J. Tissen, *Stud. Conservat.* **1967**, *12*, 41-55.
[4] M. José-Yacaman, L. Rendon, J. Arenas, M.C. Serra Puche, *Science* **1996**, *273*, 223-224.
[5] G. Chiari, R. Giustetto, G. Ricchiardi, *Eur. J. Mineral.* **2003**, *15*, 21-33.
[6] L.A. Polette-Niewold, S.F. Manciu, B. Torres, M. Alvarado Jr., R.R Chianelli, *J. Inorg. Biochem.* **2007**, *101*, 1958.
[7] B. Hubbard, W. Kuang, A. Moser, G.A. Facey, C. Detellier, *Clays Clay Minerals* **2003**, *51*, 318-326.
[8] A. Tilocca, E. Fois, *J. Phys. Chem. C* **2009**, 113, 8683-8687.
[9] F. S. Manciu, A. Ramirez, W. Durrer, J. Govani, R. R. Chianelli, *J. Raman Spectrosc.* **2008**, *39*, 1257- 1261.
[10] R. Giustetto, F X. Llabres i Xamena, G. Ricchiardi, S. Bordiga, A. Damin, R. Gobetto, M.R. Chierotti. *J. Phys. Chem. B* **2005,** *109*, 19360-19368.
[11] M. Sánchez del Río, P. Martinetto, C. Reyes-Valerio, E. Dooryhée, M. Suárez, *Archaeometry* **2006,** *48*, 115-130.

[12] C. Miliani, A. Romani, G. Favaro, *Spectrochimica Acta Part A* **1998**, *54*, 581–588.
[13] D. Reinen, P. Köhl, C. Müller, *Z. Anorg. Allg. Chem.* **2003**, *630*, 97–103.
[14] M. Leona, F. Casadio, M. Bacci, M. Picollo, *JAIC*. **2004**, *43*, 39-54.
[15] G. Chiari, R. Giustetto, J. Druzik, E. Doehne, G. Ricchiardi, *Appl. Phys. A* **2008**, *90*, 3–7.
[16] E. Fois, E. Gamba, A. Tilocca, *Microporous Mesoporous Mater.* **2003,** *57*, 263-272.
[17] M. Sánchez del Río, E. Boccaleri, M. Milanesio, G. Croce, W. van Beek, C. Tsiantos, G. D. Chyssikos, V. Gionis, G. H. Kacandes, M. Suárez, E. García-Romero, *J. Mater. Sci.* **2009**, 44, 5524–5536.
[18] J.E. Post, P.J. Heaney, *American Mineralogist* **2008**, *93*, 667-675.
[19] S. Ovarlez, F. Giulieri, A.M. Chaze, F. Delamare, J. Raya, J. Hirschinger, *Chem. Eur. J.* **2009,** *15(42)*, 11326-11332.
[20] M. E. Fernández, J. A. Ascencio, D. Mendoza-Anaya, V. Rodríguez Lugo, M. José-Yacamán, *J. Mat. Sci.* **1999**, *34*, 5243-5255.
[21] F. A. Carrey et R. J. Sundberg, *Chimie organique avancée (tomes 1 et 2)*, De Boeck, **2000**.
[22] H. Van Koningsveld, F. Tuinstra, H. Van Bekkum, J.C. Jansen, *Acta Crystallogr. B* **1989**, *45*, 423-431.
[23] P. Gómez-Romero, C. Sanchez, *New J. Chem.* **2005**, *29*, 57-58.
[24] J.L. Guth, H. Kessler, R. Weg, in *Proceedings of the Seventh International Zeolite Conference*, Vol. 137, Elsevier, Amsterdam, **1986**.
[25] J.F. Bérar, N. Boudet, S. Arnaud and B. Caillot, http://www.esrf.eu/UsersAndScience/Experiments/CRG/BM02 (2010).
[26] J. Rodríguez-Carvajal. *Physica B* **1993,** *192*, 55-69.
[27] V. Petricek, M. Dusek and L. Palatinus (2006). Jana2006. The crystallographic computing system. Institute of Physics, Praha, Czech Republic.
[28] W. Smith, T.R. Forester, I.T. Todorov, *The DL_POLY_2.0 User Manual*, SFTC Daresbury Laboratory, Warrington WA4 4AD, United Kingdom, **2009**.
[29] M. Matsui, *Physics and Chemistry of Minerals* **1996**, *23*, 345-353.
[30] H.J.C. Berendsen, J.R. Grigera, T.P. Straatsma, *J. Phys. Chem.* **1987,** *91*, 6269-6271.
[31] W.L. Jorgensen, D.S. Maxwell, J. Tirado-Rives, *J. Am. Chem. Soc.* **1996,** *118*, 11225-11236.
[32] D. Wolf, P. Keblinski, S. Phillpot, J. Eggebrecht, *J. Chem. Phys.* **1999,** *110*, 8254-8282.